\documentclass{ecai} 
\usepackage{setspace}
\usepackage{latexsym}
\usepackage{amssymb}
\usepackage{amsmath}
\usepackage{amsthm}
\usepackage{booktabs}
\usepackage{enumitem}
\usepackage{graphicx}
\usepackage{color}
\usepackage{xcolor}
\usepackage{multirow}
\graphicspath{ {./img/} }
\usepackage{algorithm}
\usepackage{algpseudocode}
\usepackage{amsmath} % 用于矩阵运算符号说明
\newcommand{\BibTeX}{B\kern-.05em{\sc i\kern-.025em b}\kern-.08em\TeX}
\definecolor{mycommentcolor}{HTML}{418080}

\algrenewcomment[1]{\hfill\small\textcolor{mycommentcolor}{#1}}
\usepackage[utf8]{inputenc}
\begin{document}

%%%%%%%%%%%%%%%%%%%%%%%%%%%%%%%%%%%%%%%%%%%%%%%%%%%%%%%%%%%%%%%%%%%%%%%%

\begin{frontmatter}

%%% Use this command to specify your submission number.
%%% In doubleblind mode, it will be printed on the first page.

\paperid{0527} 

%%% Use this command to specify the title of your paper.

% \title{PMQ: Introducing Patient Memory Queue to Patient Contrastive Learning For Electrocardiogram}

% A potential title    
\title{Enhancing Contrastive Learning-based Electrocardiogram Pretrained Model with Patient Memory Queue}   

%%% Use this combinations of commands to specify all authors of your 
%%% paper. Use \fnms{} and \snm{} to indicate everyone's first names 
%%% and surname. This will help the publisher with indexing the 
%%% proceedings. Please use a reasonable approximation in case your 
%%% name does not neatly split into "first names" and "surname".
%%% Specifying your ORCID digital identifier is optional. 
%%% Use the \thanks{} command to indicate one or more corresponding 
%%% authors and their email address(es). If so desired, you can specify
%%% author contributions using the \footnote{} command.

\author[A]{\fnms{Xiaoyu}~\snm{Sun}\footnote{Equal contribution.}}
\author[A]{\fnms{Yang}~\snm{Yang}\footnotemark}
% \author[A]{\fnms{Li}~\snm{Lin}}
\author[A]{\fnms{Xunde}~\snm{Dong}\thanks{Corresponding Author. Email: audxd@scut.edu.cn}}
% \author[A]{\fnms{First}~\snm{Author}\orcid{....-....-....-....}\thanks{Corresponding Author. Email: somename@university.edu.}\footnote{Equal contribution.}}
% \author[B]{\fnms{Second}~\snm{Author}\orcid{....-....-....-....}\footnotemark}
% \author[B,C]{\fnms{Third}~\snm{Author}\orcid{....-....-....-....}} 

\address[A]{School of Automation Science and Engineering, South China University of Technology, Guangzhou, China}
% \address[B]{Short Affiliation of Second Author and Third Author}
% \address[C]{Short Alternate Affiliation of Third Author}

%%% Use this environment to include an abstract of your paper.

\begin{abstract}
% As the volume of unlabeled medical series data continues to increase dramatically, developing a pre-training method to effectively harness this data is crucial.
% Contrastive learning is a self-supervised pre-training method that encourages representations derived from the same instance to be similar.
In the field of automatic Electrocardiogram (ECG) diagnosis, due to the relatively limited amount of labeled data, how to build a robust ECG pretrained model based on unlabeled data is a key area of focus for researchers.
Recent advancements in contrastive learning-based ECG pretrained models highlight the potential of exploiting the additional patient-level self-supervisory signals inherent in ECG.
They are referred to as patient contrastive learning.
Its rationale is that multiple physical recordings from the same patient may share commonalities, termed patient consistency, so redefining positive and negative pairs in contrastive learning as intra-patient and inter-patient samples provides more shared context to learn an effective representation.
However, these methods still fail to efficiently exploit patient consistency due to the insufficient amount of intra-inter patient samples existing in a batch.
Hence, we propose a contrastive learning-based ECG pretrained model enhanced by the \textbf{P}atient \textbf{M}emory \textbf{Q}ueue (PMQ), which incorporates a large patient memory queue to mitigate model degeneration that can arise from insufficient intra-inter patient samples.
In order to further enhance the performance of the pretrained model, we introduce two extra data augmentation methods to provide more perspectives of positive and negative pairs for pretraining.
Extensive experiments were conducted on three public datasets with three different data ratios. The experimental results show that the comprehensive performance of our method outperforms previous contrastive learning methods and exhibits greater robustness in scenarios with limited labeled data.
The code is available at https://github.com/3hiuwoo/PMQ.
% We show that our method outperforms previous patient contrastive learning methods in average across multiple ECG downstream datasets and exhibits greater robustness in scenarios with limited labeled data.
% This improvement signifies that our approach facilitates the learning of more generalizable representations by optimizing on a broader range of positive and negative samples generated within the patient-level context.
\end{abstract}

\end{frontmatter}

%%%%%%%%%%%%%%%%%%%%%%%%%%%%%%%%%%%%%%%%%%%%%%%%%%%%%%%%%%%%%%%%%%%%%%%%

\section{Introduction}

The electrocardiogram (ECG) is a non-invasive method for measuring the heart's electrical activity and has gained increasing importance for detecting and diagnosing cardiac diseases.
Numerous deep learning methods have been introduced to learn the intricate patterns inherent in the complex periodic rhythms of ECG \cite{surveydlecg}.
However, due to the challenge of obtaining high-quality manual labels of ECG, which is labor-intensive for physicians, these models are hindered by data scarcity.
Self-supervised learning (SSL) \cite{bert, moco, mae, rotation, colorization, timesiam} offers a way to address the problem by taking advantage of the extensive unlabeled data available on the Internet.
As a way of SSL, contrastive learning \cite{surveycl} has demonstrated significant advantages in computer vision and has attracted widespread attention across various domains.
The core idea of contrastive learning is the instance discrimination pretext task \cite{memorybank} that compels the model to learn similar representations for positive samples augmented from the same data instance and dissimilar representations for negative samples from different data instances.

Motivated by its success, the research community has adopted and further developed contrastive learning for ECG analysis.
Among them, a line of previous work leverages the additional data level of the ECG series: patient level \cite{clocs, pclr, comet}.
These methods extend contrastive learning by leveraging considerably more positive samples under intra-patient contexts to learn a more generalizable representation for ECG data.
We refer to them as patient contrastive learning, as they utilize the fact that multiple physical recording instances can share meaningful context such as periodic cardiac patterns in ECG if derived from the same patient \cite{clocs}.
Benefiting from the additional data level, these methods have demonstrated several advantages over contrastive learning methods designed for instance level, such as less dependency on augmentation design \cite{pclr} and patient-specific representation \cite{clocs}.

Existing methods, however, are limited in their ability to fully capture patient-level shared context.
Specifically, during training on large pretraining datasets, only a mini-batch of data is sampled at each iteration.
As a result, data from the same patient are unlikely to appear within the same batch, leading to a scarcity of intra-patient positive pairs.
This limitation reduces the approach to standard instance-level contrastive learning \cite{simclr}.
To alleviate this, methods like COMET \cite{comet} have elaborated on warping the random batch sampler to ensure the quantity of positive samples.
Rather than thoroughly shuffling the data, it first groups samples from the same trial into sets, and then shuffles the order of samples within each set.
In the end, all patient sets are sorted while preserving the internal order of samples.
As a result, all samples from the same trial will be included in the same batch.
Because samples from the same trial are naturally from the same patient, thus the number of positive samples is increased.
However, due to the incomplete shuffling and constrained batch size, each batch still contains too few patients to provide diverse inter- and intra-patient samples.

\begin{figure*}[ht]
    \centering
    \includegraphics[width=0.95\linewidth, trim=25 180 20 90, clip]{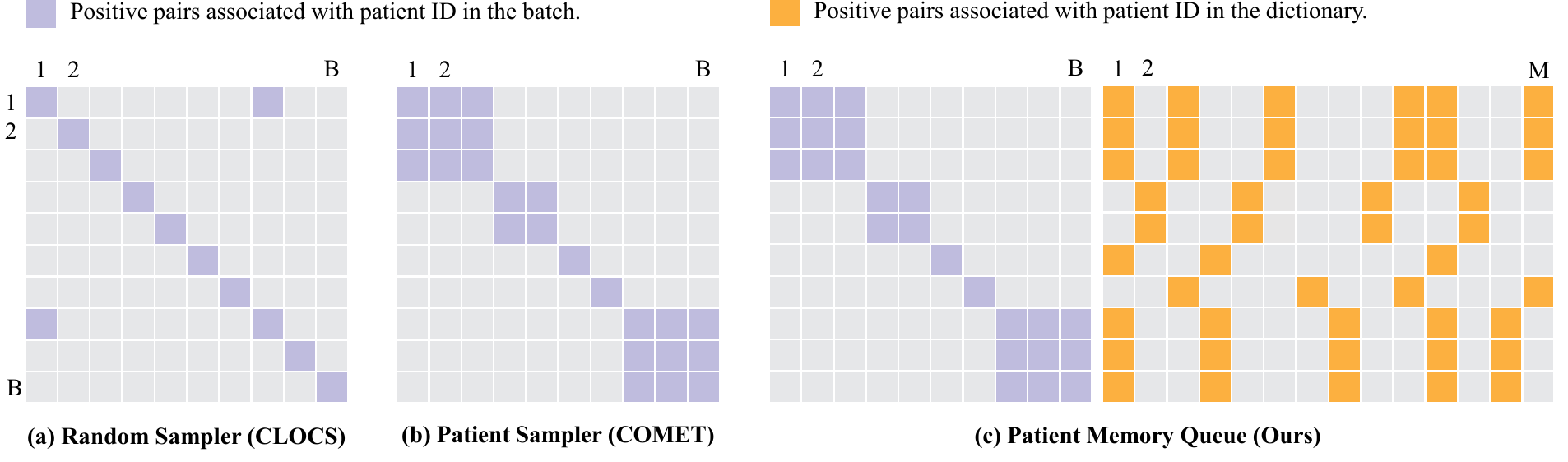}
    \caption{\textbf{Difference between ours method (PMQ) and previous methods.}
    Different similarity matrices used for computing contrastive loss between two augmented views of a mini-batch of size $B$ with corresponding patient IDs $\{\,p_1,\, \ldots,\, p_B \,\}$.
    Each element at row $i$ and column $j$ represents the cosine similarity between the first augmented view of the $i$-th sample and the second augmented view of the $j$-th sample. 
    % Diagonal elements always correspond to positive pairs, as they originate from the same sample.
    (a) Early patient contrastive learning methods, such as CLOCS \cite{clocs}, randomly shuffle the dataset and sample a mini-batch irrespective of patient identity. In addition to diagonal elements, off-diagonal entries $(\,i,\, j\,)$ are also treated as positives if $p_i = p_j$.
    (b) Approaches like COMET \cite{comet} apply hierarchical shuffling to retain trial-level grouping, sampling mini-batches sequentially.
    This increases the likelihood of neighboring samples sharing the same patient ID within a batch.
    (c) In contrast to prior methods, we incorporate an external patient memory queue containing $M$ representations and their associated patient IDs.
    At each iteration, an additional similarity matrix is computed between the mini-batch and the patient memory queue.
    % This enables the inclusion of a substantially larger number of positive samples during training, thereby enhancing the model’s ability to capture patient-level characteristics more effectively.
    }
    \label{fig:improvement}
\end{figure*}

Motivated by this gap, we introduce a patient memory queue as an auxiliary module to the end-to-end paradigm.
An shown in Figure~\ref{fig:improvement},
by storing a substantial number of prior representations from different patients, the queue ensures an adequate quantity of intra-inter patient samples \cite{hardneg, awasthiMoreNegativeSamples2022}, enhancing the patient contrastive learning model by effectively exploiting patient consistency.
% Note that while enlarging the batch size assumably ensures the quantity of samples, using a large batch size often encounters optimization problems \cite{moco} and never reaches the same size as the patient memory dictionary.

Our contributions are summarized as the following:
\begin{enumerate}
    \item We proposed a patient contrastive learning method that incorporates a dynamic \textbf{P}atient \textbf{M}emory \textbf{Q}ueue, entitiled PMQ, which stores a great number of positive and negative samples under patient context. This enables the pretraining process to maximize the exploration of context information in ECG series.
    \item We introduce extra data augmentation techniques: timestamp masking and frequency masking. By sequentially superimposing these two data augmentation methods on the basis of the neighbor view, more instance discriminative representations for perturbations can be learned. 
    \item We conduct extensive experiments on three public datasets across three different labeled data ratios. The results demonstrate that our method consistently outperforms existing patient-level and general contrastive learning approaches, particularly in scenarios with limited labeled data, where it exhibits enhanced robustness. 
    % Furthermore, our method achieves performance that is comparable to—or even surpasses that of approaches leveraging additional ECG textual reports and large language models, despite relying solely on raw ECG data.
\end{enumerate}

\section{Related work}

\paragraph{Contrastive learning.}
%A wide range of works about contrastive learning have followed the instance discrimination pretext task, where two augmented samples that originate from the same instance (an image, a time series, etc.) are regarded as a positive pair, and otherwise, negative.
Contrastive learning designs the instance discrimination pretext task \cite{memorybank} to learn representations that are similar if they come from augmented views \cite{moco, simclr, byol} of the same data; otherwise, they are dissimilar.
The time series community also embraces the contrastive learning method to learn the transferable representation \cite{surveysslts}.
Because the pattern varies in time series from different domains such as finance, industry, healthcare, etc., previous studies have exploited different shared contexts in the time series.
Typical time series contrastive learning has explored the transformation consistency by various augmentation methods \cite{tstcc, mtts, dtw}.
TS2Vec \cite{ts2vec} performs multi-scale contrastive learning to learn a fine-grained contextual representation.
There are also works \cite{cost, btsf, tfc, timesurl, fusion} learn the generalizable representation by explicitly leveraging the frequency information through masking \cite{tfc} or mixing \cite{timesurl} the components of the spectra, time-frequency fusion \cite{cost, btsf, fusion}, and hierarchical learning \cite{hirarc}.
Since the ECG belongs to the time series, existing methods entailed in the general time series can be applied to ECG self-supervised learning as well, but they fall short of leveraging the extra context inherent in the medical time series.
Our works focus on the additional patient-level context introduced by ECG series.

\paragraph{Self-supervised learning for ECG.}

A variety of self-supervised learning (SSL) methods have been tailored for electrocardiogram (ECG) signals, harnessing their intrinsic structure and temporal dependencies.
Techniques such as predicting missing samples \cite{maefe} and utilizing augmentation classes \cite{mtssl} exemplify such adaptations.
Contrastive learning, previously outlined, is also a significant SSL strategy within the ECG domain, emphasizing the spatial and temporal attributes to create effective positive and negative pairs \cite{clocs, hu2023spatiotemporal, chenTemporalSpatialSelf2025}.
Additionally, some studies have ventured into utilizing associated text reports to bolster ECG representation learning \cite{etp, eclip}.
Despite these advancements, many methodologies overlook the potential of leveraging supplementary contextual information, particularly at the patient level.
CLOCS \cite{clocs} addresses this gap by introducing a patient contrastive learning mechanism, reformulating positive pairs as augmented views from the same patient.
Additionally, it implements temporal and spatial augmentations specifically tailored for ECG data.
PCLR \cite{pclr} utilizes patient consistency by pairing two independent samples from the same patient, underscoring the significance of harnessing patient-level signals.
COMET \cite{comet}, integrating TS2Vec, advances the concept through a hierarchical contrastive learning framework, extending from sample and instance levels to trial and patient levels.
However, it encounters optimization challenges stemming from conflicts in selecting positive pairs across different levels.
While patient contrastive learning strategies benefit from patient contexts, they do not fully capitalize on this resource.
Owing to the batch sampling mechanism, patient-level positive pairs are often sparse in each training iteration, meaning that patient consistency is rarely leveraged in the practical training process.
Our work aims to explore the potential of fully utilizing the shared patient-level context by increasing the number of positive pair.

\section{Methods}

% In this section, we first outline the problem formulation with a set of notations. Secondly, we give a unified modeling of the patient discrimination pretext task. Finally, we present the architecture of our method.
\begin{figure*}[ht]
    \centering
    \includegraphics[width=\linewidth, trim=70 175 100 180, clip]{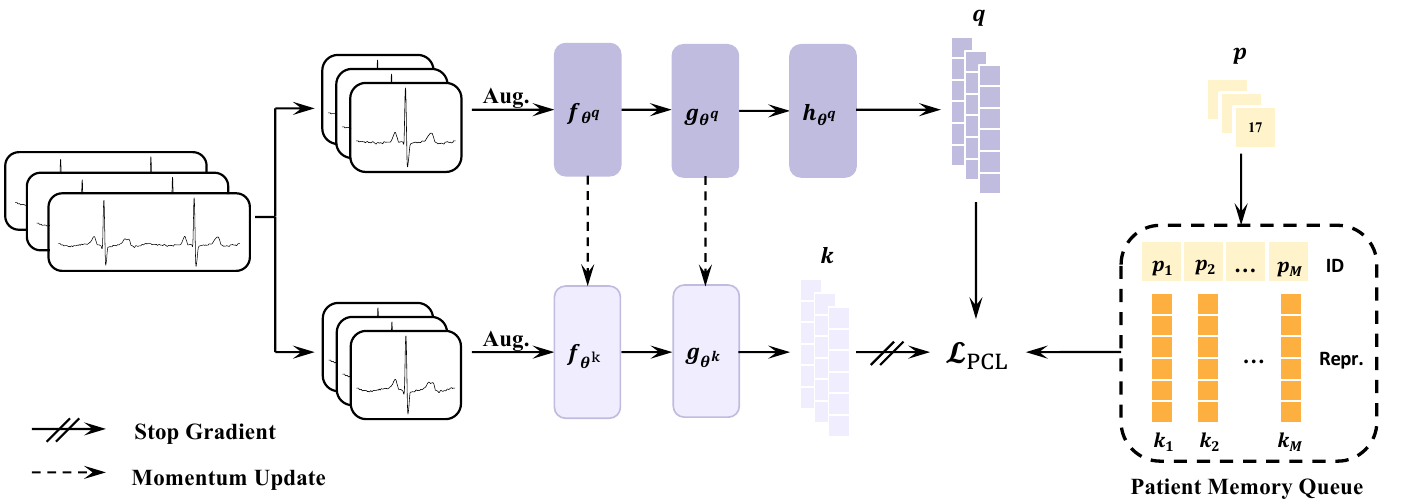}
    \caption{\textbf{Overview of PMQ approach.},
    The patient memory queue stores previous representations (Repr.) with their patient ID simultaneously. When training, A mini-batch of samples and their associated patient IDs $p$ is sampled. Each sample is augmented into two views and passed through the query encoder $f{\theta^q}$ and the key encoder $f_{\theta^k}$, respectively.
    In our implementation, we first sample two neighboring segments, then apply temporal and frequency masking to generate two distinct augmented views.
    The output of each encoder is subsequently processed by projection heads, with an additional prediction head applied only to the query branch, yielding the final query and key representations.
    The patient contrastive loss is computed by cosine similarity matrices: one between the query and key representations, and another between the query representations and the memory queue $\mathcal{Q}$.
    Positive pairs are identified based on matching patient IDs between $p$ itself for the first matrix and between $p$ and $\mathcal{Q}$ for the second matrix.
    After computing the loss, parameters in the query branch (upper path) are updated via backpropagation. The key branch (lower path) is updated using a moving average of the query branch parameters, excluding the prediction head.
    }
    \label{fig:pipeline}
\end{figure*}

\subsection{Preliminary}
\label{sec:prob}

Let an unlabeled ECG dataset consist of $P$ patients, where each patient has one or more trials.
Following \cite{comet} we segment all trials into heartbeat-level samples with equal length, producing the pretraining dataset: $\mathcal{D}_{\text{pretrain}} \in \mathbb{R}^{N \times S \times L}$, where $N$ is the total number of samples, $S$ is the length of each sample, $L$ is the number of leads.
We assign each sample a patient ID $p \in \{\,0,\, 1,\, \ldots,\, p-1 \,\}$ to indicate its source patient.
We produce several downstream finetuning datasets $\mathcal{D}_{finetune}$ in the same manner, where each sample has a label $y \in \{\, 0,\, 1,\, \ldots,\, C-1 \,\}$ representing the cardiac rhythm.

We train an encoder $f_{\theta}: \mathbb{R}^{S \times L} \rightarrow \mathbb{R}^{K}$ parameterized by $\theta$ on $\mathcal{D}_{pretrain}$ by self-supervised representation learning.
Then, we transfer it to $\mathcal{D}_{\text{finetune}}$ for downstream ECG classification tasks.
By exploiting patient consistency, our goal is to learn a representation benefiting the downstream task performance and obtaining stronger robustness to less labeled data.

% \subsection{Patient Discriminative Modeling}
\subsection{Patient Contrastive Learning with Memory Queue}
\label{pmq}
% We assume that all trials from the same patient were collected in the same setting and each is normal or has a disease following \cite{clocs, comet} to ensure patient consistency.
% With this caveat, the patient discrimination pretext task refers to learning representations from the same patient sharing the most commonalities.

% Incorporated with contrastive learning \cite{lecun}, we consider samples originating from the same patient as positive pairs and otherwise as negative pairs.
% Different from the instance discrimination \cite{memorybank, moco, simclr}, multiple positive pairs can be retrieved since a patient has numerous samples.
Inspired by MoCo \cite{moco}, we reformulate patient contrastive learning as a one-to-many dictionary look-up problem, where the key is the patient ID associated with a sample, and the value is the encoded representation of that sample. 
Existing patient contrastive learning methods can be interpreted as maintaining a dictionary limited to the current mini-batch, where all contents are discarded after each iteration. 
As a result, the number of positive intra-patient and negative inter-patient samples is constrained by the batch size, limiting the diversity of patient-level information available during training.
We hypothesize that expanding the dictionary size can enhance patient contrastive learning by incorporating more diverse and informative patient representations. 
To address this, we propose the patient memory queue to serve as the dictionary, which decouples the number of patient-positive and patient-negative samples from the mini-batch size.
This design enables the construction of a substantially larger and more persistent patient memory dictionary.
While previous MoCo-based approaches \cite{moco, clecg} also employ memory queues to increase the number of negative samples, they are not specifically tailored to patient contrastive learning and overlook positive intra-patient samples. 
In contrast, PMQ explicitly models both positive intra-patient and negative inter-patient samples, both of which are crucial for effective patient contrastive learning.
The whole architecture of our method is depicted in Figure~\ref{fig:pipeline}.

Specifically, for an input sample $x$ with patient ID $p$, we generate two augmented views $x_q, x_k$.
Then, we obtain query representation $q=f^q(x_q)$ and key representation $k=f^k(x_k)$ from the query encoder and key encoder respectively.
The encoders are to be described in Section~\ref{sec:encoder}.
We maintain a patient memory queue $\mathcal{Q}$ as the dictionary with size $M$: $\{\,(\,p_0,\, k_0\,),\, (\,p_1,\, k_1\,),\, \ldots,\, (\,p_{M-1},\, k_{M-1}\,)\,\}$,
where previous encoded key representations are saved as values and their patient IDs as keys.
To compute patient contrastive loss, we firstly enqueue the current key representation and patient ID to update the patient memory queue.
Then, we retrieve all representations $k_i$ from the patient memory queue with index $i:p_i=p_q$.
The loss function serves as the self-supervision metric which yields a high value if the query patient representation is similar to representations from the same patient in the queue.

In summary, referring to InfoNCE loss \cite{cpc}, we define the loss of Patient Contrastive Learning as follows:
\begin{eqnarray}
\label{eq:loss1}
\mathcal{L}_{\text{PCL}} = 2\tau\mathbb{E}_{x_i \in \mathcal{B}} \left[\,
\mathbb{E}_{k^+ \in P_i^+} \left[ -\log \frac{\exp(q_i \cdot k^+ / \tau)}
{\sum_{k \in \mathcal{Q}} \exp(q_i \cdot k / \tau)} \right] \,
\right]
\end{eqnarray}
where $P_i^+=\{\,k^+ \mid p^+ = p_i\,\}$ denotes all positive keys associated with patient $p_i$ retrieved from the patient memory queue, $\tau \in [\,0,\, 1\,]$ is a temperature hyper-parameter, and $\mathcal{B}$ is the mini-batch.
We scale the loss by $2\tau$ as it makes the model less sensitive to the $\tau$ value \cite{byol}.
Note that all query and key representations are $L_2$-normalized to enable cosine similarity computation and to stabilize the training process.
Additionally, all samples stored in the memory queue are utilized as negative samples.

After the loss computation, we dequeue the earliest batch of representations and patient IDs to keep the patient memory queue up to date in a sense that outdated representations might violate patient consistency because the encoder is evolving \cite{moco}.

% Previous patient contrastive learning methods can be interpreted as maintaining a dictionary equivalent to the batch size.
% The encoded representations are flushed immediately after each iteration; thus, the number of positive and negative samples is limited by the batch size. 
% By introducing the patient memory dictionary, we make up for previous patient contrastive learning methods by decoupling the patient positive and negative pair quantity from the mini-batch size, enhancing the model's performance.

\subsection{Momentum Update}
\label{sec:encoder}

% The patient memory queue consists of a sub-queue of preceding representations and a sub-queue of corresponding patient IDs.
% We term them as patient representation queue $\mathcal{Q}_{\text{repr}} \in \mathbb{R}^{M \times K}$ and patient ID queue $\mathcal{Q}_{\text{id}} \in \mathbb{R}^{M}$.
% The patient memory dictionary can be represented as $\{\,(\,\mathcal{Q}^i_{id},\, \mathcal{Q}^i_{repr}\,)\,\},i \in \{\,0,\, \ldots,\, M-1\,\}$.
% In the training process, the encoded mini-batch and the corresponding patient IDs are enqueued to two queues respectively for updating the dictionary.

The patient memory queue is required to update progressively for preventing the model from being confused by the rapid changes between successively enqueued representations \cite{moco}.
To implement this, we train the query encoder $f_{\theta^q}$ referring to $f_{\theta}$ in Section~\ref{sec:prob} by backpropagation, and adopt a momentum key encoder $f_{\theta^k}$ following \cite{moco, byol}, which is updated smoothly by moving average:
\begin{eqnarray}
    \theta_k \leftarrow m \theta_k + (1 - m)\theta_q
\end{eqnarray}
where $\theta_q$ and $\theta_k$ denote the parameters of $f^q$ and $f^k$. $m \in [\,0,\, 1\,)$ is another hyper-parameter that controls the momentum encoder evolving speed.

We attach projection heads $g_{\theta^q}$ and $g_{\theta^k}$ to the query and key encoders, respectively, to project the learned representations into the loss space, following the design in \cite{simclr}.
The projection head associated with the momentum-updated key encoder is synchronized via a moving average of the parameters from its counterpart on the query side.
Additionally, we introduce a prediction head $h_{\theta^q}$ exclusively on top of the query projection head, as inspired by \cite{byol}.
The query representation $q$ is the output of the prediction head and the key representation $k$ is the output of the momentum-updated projection head.
This asymmetric architecture is intended to mitigate the discrepancy in update dynamics between the query and key encoders.

As a result, we build a large and dynamic patient memory queue that keeps track of the latest progressive representations and preserves patient consistency, enabling the full exploitation of effective representation.

\begin{figure}[ht]
    \centering
    \includegraphics[width=\linewidth, trim=10 30 500 10, clip]{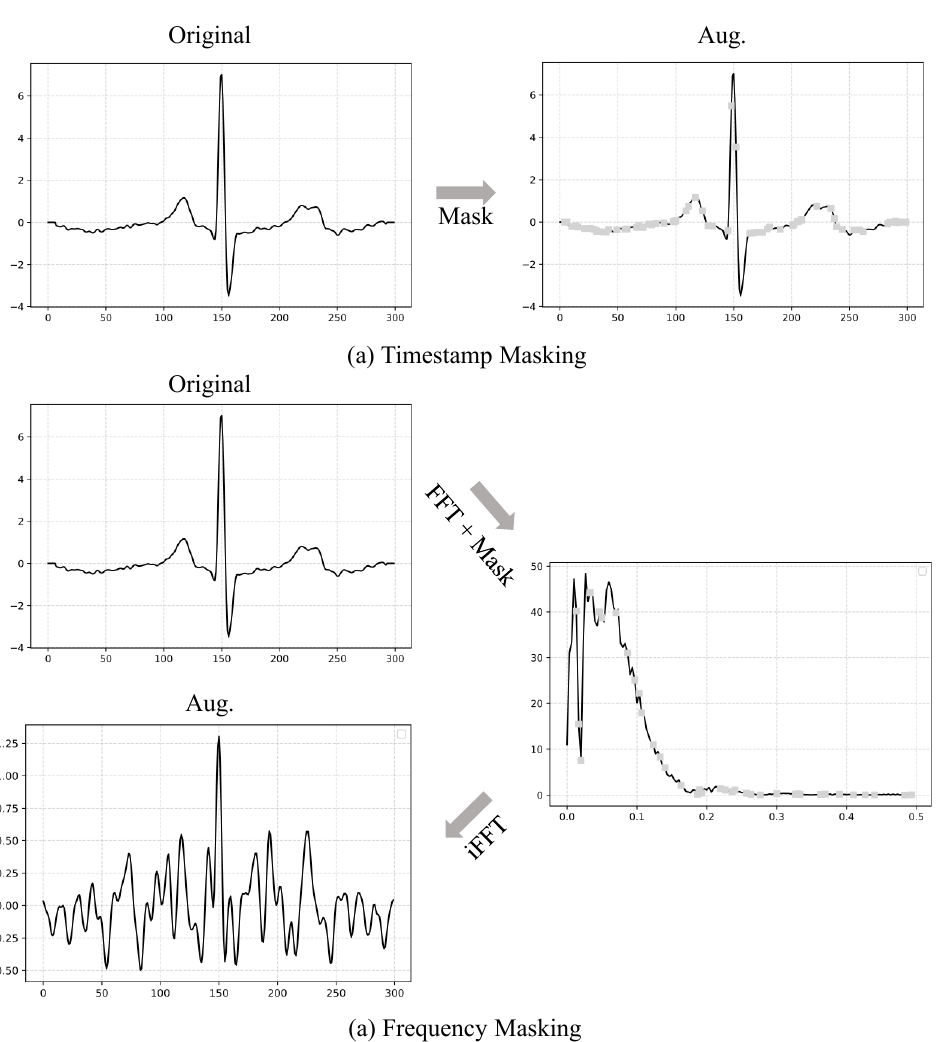}
    \caption{\textbf{Visualization of the time and frequency masking.} For simplicity, only a single lead of ECG without projection is masked.}
    \label{fig:aug}
\end{figure}

\subsection{Data Augmentations}
With the presence of numerous positive pairs naturally existing at the patient level, a profound advancement in patient contrastive learning is saving the labor of designing intricate data augmentation techniques to learn instance-level representations \cite{timesiam, comet, pclr}.

In addition, we elaborate on plug-and-play data augmentation methods to introduce more variations into the training stage, aiming to learn a more robust and discriminative representation with the supplement of instance-level features.\footnote{Meanwhile, if no perturbations are applied, computing the cosine similarity between a positive pair originating from the same sample can be trivial.}

We employed three straightforward yet effective stochastic data augmentation techniques to leverage the temporal, spatial, and spectral characteristics of ECG.
These techniques are applied sequentially to the input ECG sample to generate augmented views.
% As the primary focus of this work is not on developing novel augmentation techniques, we adopt several simple yet representative augmentations to demonstrate the effectiveness of incorporating additional augmentation strategies.
% Specifically, we explore transformation invariance across both time and frequency domains, which is a property broadly preserved in time series data.\cite{tstcc, clocs, tfc, cost, btsf}
% We do research on the individual effect of each augmentation technique in Section~\ref{sec:ablation}

\paragraph{Temporal Neighboring.}
Instead of randomly sampling a segment, we sample two neighboring segments as the query ($x_q$) and key ($x_k$), which shown in the Figure~\ref{fig:pipeline}
This encourages the model to learn richer temporal dependencies by leveraging the assumption that temporally adjacent segments share higher mutual information \cite{clocs, naGuidingMaskedRepresentation2024, chenTemporalSpatialSelf2025, hu2023spatiotemporal, wangAdversarialSpatiotemporalContrastive2024, tnc}.
\begin{align*}
x &= \text{Segment}(t,\, t + 2\Delta t)\\
x_q &= \text{Segment}(t,\, t + \Delta t) \\
x_k &= \text{Segment}(t + \Delta t,\, t + 2\Delta t)
\end{align*}
where $x$ represents a randomly sampled segment.

\paragraph{Timestamp Masking.}
Following \cite{ts2vec, comet}, each input segment is first projected into a higher-dimensional embedding space, where the input dimension corresponds to the number of leads.
A binary mask $M \in \{\,0,\, 1\,\}$ is then applied independently to each timestamp with probability $p$.
Performing masking in the projected space avoids unintentionally masking zero values in the raw input and introduces discrete perturbations that serve as effective augmentations \cite{ts2vec}.
\begin{align*}
E &= f(x) \in \mathbb{R}^{S \times D} \\
M &\in \{0, 1\}^{L \times 1} \\
M_i &\sim \text{Bernoulli}(p) \quad \forall i \in \{1, 2, \ldots, N\} \\
E' &= E \odot M
\end{align*}
where $ E $ is the projected embedding matrix, $ D $ is the dimensionality of the higher-dimensional space, and $ f $ is the projection function. $ E' $ is the masked embedding matrix, and $ \odot $ denotes element-wise multiplication. The visualization of the timestamp masking is shown in Figure~\ref{fig:aug} (a).

\paragraph{Frequency Masking.}
Prior to timestamp masking, we transform the projected input into the frequency domain using the Fast Fourier Transform (FFT).
A small subset of frequency components is randomly selected, and their amplitudes are set to zero \cite{tfc}.
Then, the modified spectra is transformed back to the time domain using the inverse FFT (iFFT).
This process introduces smooth, frequency-specific perturbations, serving as a complementary form of continuous augmentation.
\begin{align*}
E &= f(x) \in \mathbb{R}^{S \times D} \\
F &= \text{FFT}(E) \in \mathbb{C}^{\lfloor \frac{S}{2} \rfloor + 1 \times D} \\
F_{j} &= 0 \quad \text{for randomly selected components } j \in \mathbb{R}^D \\
E' &= \text{iFFT}(F) \in \mathbb{R}^{S \times D}
\end{align*}
where $ F $ is the frequency spectrum of the embedded input, $ j $ denotes the frequency index for all leads. The visualization of the frequency masking is shown in Figure~\ref{fig:aug} (b).
%\paragraph{Frequency excahnge.} proposed by timesurl

\section{Experiments}

\begin{table*}[ht]
\centering
\begin{spacing}{1.1} % 调整行距
\caption{\textbf{The experimental results of various pre-training methods after fine-tuning on different downstream datasets/different data ratios.} Among them, "Random" means that the weights of the pre-trained model are randomly initialized. }
\label{tab:result}
\resizebox{\linewidth}{!}{
\begin{tabular}{ccccccccccc} 
\hline\hline
\multicolumn{11}{c}{\rule{0pt}{10pt}\fontsize{8.5}{10}\selectfont  \textbf{Data ratio: 30\%}}                                                                                                                \\ 
\hline
\multirow{2}{*}{Method} & \multicolumn{3}{c}{PTB-XL}     & \multicolumn{3}{c}{Chapman}    & \multicolumn{3}{c}{CPSC2018}   & \multirow{2}{*}{Overall}  \\ 
\cline{2-10}
                        & F1       & AUROC    & ACC      & F1       & AUROC    & ACC      & F1       & AUROC    & ACC      &                           \\ 
\hline
Random                  & 56.0\textcolor{gray}{\tiny$\pm$0.8} & 84.0\textcolor{gray}{\tiny$\pm$0.9} & 69.9\textcolor{gray}{\tiny$\pm$0.5} & 86.6\textcolor{gray}{\tiny$\pm$1.0} & 95.9\textcolor{gray}{\tiny$\pm$3.0} & 85.6\textcolor{gray}{\tiny$\pm$0.8} & 59.7\textcolor{gray}{\tiny$\pm$1.1} & 90.4\textcolor{gray}{\tiny$\pm$0.6} & 63.5\textcolor{gray}{\tiny$\pm$0.6} & 76.8                      \\ 
\hline
MOCO                    & 53.5\textcolor{gray}{\tiny$\pm$0.3} & 83.7\textcolor{gray}{\tiny$\pm$0.8} & 70.8\textcolor{gray}{\tiny$\pm$1.1} & 84.5\textcolor{gray}{\tiny$\pm$0.5} & 95.5\textcolor{gray}{\tiny$\pm$0.5} & 82.4\textcolor{gray}{\tiny$\pm$0.8} & 64.0\textcolor{gray}{\tiny$\pm$0.5} & 91.2\textcolor{gray}{\tiny$\pm$0.2} & 67.2\textcolor{gray}{\tiny$\pm$0.7} & 77.0                        \\
BYOL                    & 54.9\textcolor{gray}{\tiny$\pm$0.4} & 84.4\textcolor{gray}{\tiny$\pm$0.7} & 71.1\textcolor{gray}{\tiny$\pm$0.5} & 86.3\textcolor{gray}{\tiny$\pm$1.5} & 96.5\textcolor{gray}{\tiny$\pm$0.4} & 84.2\textcolor{gray}{\tiny$\pm$2.0} & 63.3\textcolor{gray}{\tiny$\pm$1.2} & 91.2\textcolor{gray}{\tiny$\pm$0.3} & 64.3\textcolor{gray}{\tiny$\pm$1.2} & 77.4                      \\
CLOCS                   & 47.3\textcolor{gray}{\tiny$\pm$0.4} & 78.5\textcolor{gray}{\tiny$\pm$0.6} & 65.9\textcolor{gray}{\tiny$\pm$0.6} & 86.1\textcolor{gray}{\tiny$\pm$0.8} & 95.3\textcolor{gray}{\tiny$\pm$0.1} & 84.9\textcolor{gray}{\tiny$\pm$1.1} & 58.7\textcolor{gray}{\tiny$\pm$0.6} & 89.5\textcolor{gray}{\tiny$\pm$0.2} & 63.9\textcolor{gray}{\tiny$\pm$0.3} & 74.5                          \\
PCLR                    & 54.2\textcolor{gray}{\tiny$\pm$0.2} & 85.1\textcolor{gray}{\tiny$\pm$0.4} & 71.1\textcolor{gray}{\tiny$\pm$0.1} & 84.6\textcolor{gray}{\tiny$\pm$0.3} & 95.6\textcolor{gray}{\tiny$\pm$0.1} & 82.5\textcolor{gray}{\tiny$\pm$0.6} & 60.5\textcolor{gray}{\tiny$\pm$1.6} & 91.5\textcolor{gray}{\tiny$\pm$0.4} & 66.9\textcolor{gray}{\tiny$\pm$0.3} & 76.9                      \\
COMET                   & 54.2\textcolor{gray}{\tiny$\pm$1.6} & 84.2\textcolor{gray}{\tiny$\pm$0.6} & 68.8\textcolor{gray}{\tiny$\pm$1.4} & 87.1\textcolor{gray}{\tiny$\pm$0.4} & 96.7\textcolor{gray}{\tiny$\pm$0.4} & \textbf{86.4}\textcolor{gray}{\tiny$\pm$0.6} & 59.7\textcolor{gray}{\tiny$\pm$1.0} & 90.6\textcolor{gray}{\tiny$\pm$0.7} & 64.4\textcolor{gray}{\tiny$\pm$1.2} & 76.9                      \\
ETP                     & 53.8\textcolor{gray}{\tiny$\pm$0.3} & 83.4\textcolor{gray}{\tiny$\pm$0.3} & 71.6\textcolor{gray}{\tiny$\pm$0.4} & 83.2\textcolor{gray}{\tiny$\pm$0.6} & 95.5\textcolor{gray}{\tiny$\pm$0.4} & 80.1\textcolor{gray}{\tiny$\pm$0.8} & 58.2\textcolor{gray}{\tiny$\pm$0.9} & 88.9\textcolor{gray}{\tiny$\pm$0.2} & 63.4\textcolor{gray}{\tiny$\pm$0.4} & 75.3                      \\ 
\hline
\textbf{Ours}                    & \textbf{56.1}\textcolor{gray}{\tiny$\pm$0.1} & \textbf{85.7}\textcolor{gray}{\tiny$\pm$0.4} & \textbf{71.7}\textcolor{gray}{\tiny$\pm$0.3} & \textbf{86.8}\textcolor{gray}{\tiny$\pm$0.9} & \textbf{96.8}\textcolor{gray}{\tiny$\pm$0.3} & 85.0\textcolor{gray}{\tiny$\pm$1.3} & \textbf{64.2}\textcolor{gray}{\tiny$\pm$0.9} & \textbf{91.7}\textcolor{gray}{\tiny$\pm$0.3} & \textbf{69.0}\textcolor{gray}{\tiny$\pm$0.6} & \textbf{78.6}                      \\ 
\hline\hline
\multicolumn{11}{c}{\rule{0pt}{10pt}\fontsize{8.5}{10}\selectfont  \textbf{Data ratio: 10\%}}                                                                                                                 \\ 
\hline
\multirow{2}{*}{Method} & \multicolumn{3}{c}{PTB-XL}     & \multicolumn{3}{c}{Chapman}    & \multicolumn{3}{c}{CPSC2018}   & \multirow{2}{*}{Overall}  \\ 
\cline{2-10}
                        & F1       & AUROC    & ACC      & F1       & AUROC    & ACC      & F1       & AUROC    & ACC      &                           \\ 
\hline
Random                  & 51.8\textcolor{gray}{\tiny$\pm$1.0} & 82.3\textcolor{gray}{\tiny$\pm$0.7} & 67.1\textcolor{gray}{\tiny$\pm$1.8} & 80.8\textcolor{gray}{\tiny$\pm$0.9} & 93.4\textcolor{gray}{\tiny$\pm$0.4} & 79.2\textcolor{gray}{\tiny$\pm$1.3} & 55.4\textcolor{gray}{\tiny$\pm$1.0} & 87.8\textcolor{gray}{\tiny$\pm$0.6} & 60.3\textcolor{gray}{\tiny$\pm$0.5} & 73.1                      \\
\hline
MOCO                    & 51.5\textcolor{gray}{\tiny$\pm$0.7} & 79.8\textcolor{gray}{\tiny$\pm$1.5} & 68.1\textcolor{gray}{\tiny$\pm$1.3} & 81.7\textcolor{gray}{\tiny$\pm$1.0} & 94.3\textcolor{gray}{\tiny$\pm$0.7} & 79.2\textcolor{gray}{\tiny$\pm$1.3} & 60.2\textcolor{gray}{\tiny$\pm$0.7} & 89.0\textcolor{gray}{\tiny$\pm$0.3} & 63.9\textcolor{gray}{\tiny$\pm$1.0} & 74.2                      \\
BYOL                    & 51.5\textcolor{gray}{\tiny$\pm$0.8} & 81.1\textcolor{gray}{\tiny$\pm$1.6} & 69.3\textcolor{gray}{\tiny$\pm$1.7} & 82.1\textcolor{gray}{\tiny$\pm$0.7} & 94.8\textcolor{gray}{\tiny$\pm$0.8} & 78.8\textcolor{gray}{\tiny$\pm$1.9} & 59.5\textcolor{gray}{\tiny$\pm$0.6} & 88.9\textcolor{gray}{\tiny$\pm$0.2} & 65.4\textcolor{gray}{\tiny$\pm$0.4} & 74.6                      \\
CLOCS                   & 45.5\textcolor{gray}{\tiny$\pm$0.5} & 76.7\textcolor{gray}{\tiny$\pm$1.2} & 64.0\textcolor{gray}{\tiny$\pm$1.6} & 81.6\textcolor{gray}{\tiny$\pm$1.0} & 93.6\textcolor{gray}{\tiny$\pm$0.2} & 79.4\textcolor{gray}{\tiny$\pm$1.5} & 57.3\textcolor{gray}{\tiny$\pm$0.3} & 88.5\textcolor{gray}{\tiny$\pm$0.3} & 61.8\textcolor{gray}{\tiny$\pm$0.1} & 72.0                           \\
PCLR                    & 52.8\textcolor{gray}{\tiny$\pm$0.3} & 83.0\textcolor{gray}{\tiny$\pm$0.7} & 70.9\textcolor{gray}{\tiny$\pm$0.8} & 80.6\textcolor{gray}{\tiny$\pm$0.7} & 93.8\textcolor{gray}{\tiny$\pm$0.3} & 77.8\textcolor{gray}{\tiny$\pm$1.1} & 59.3\textcolor{gray}{\tiny$\pm$1.0} & 89.2\textcolor{gray}{\tiny$\pm$0.2} & 63.6\textcolor{gray}{\tiny$\pm$1.2} & 74.6                      \\
COMET                   & 48.2\textcolor{gray}{\tiny$\pm$1.4} & 79.6\textcolor{gray}{\tiny$\pm$0.9} & 64.7\textcolor{gray}{\tiny$\pm$1.2} & \textbf{83.8}\textcolor{gray}{\tiny$\pm$0.6} & 94.8\textcolor{gray}{\tiny$\pm$0.8} & \textbf{82.6}\textcolor{gray}{\tiny$\pm$0.9} & 54.8\textcolor{gray}{\tiny$\pm$2.5} & 88.0\textcolor{gray}{\tiny$\pm$1.0} & 61.0\textcolor{gray}{\tiny$\pm$2.1} & 73.1                      \\
ETP                     & 52.0\textcolor{gray}{\tiny$\pm$0.6} & 80.0\textcolor{gray}{\tiny$\pm$0.3} & 70.3\textcolor{gray}{\tiny$\pm$0.6} & 78.5\textcolor{gray}{\tiny$\pm$0.8} & 93.3\textcolor{gray}{\tiny$\pm$0.4} & 74.2\textcolor{gray}{\tiny$\pm$1.1} & 57.4\textcolor{gray}{\tiny$\pm$1.0} & 88.8\textcolor{gray}{\tiny$\pm$0.2} & 64.3\textcolor{gray}{\tiny$\pm$0.8} & 73.2                      \\ 
\hline
\textbf{Ours}                    & \textbf{53.6}\textcolor{gray}{\tiny$\pm$0.3} & \textbf{83.7}\textcolor{gray}{\tiny$\pm$0.4} & \textbf{71.8}\textcolor{gray}{\tiny$\pm$0.3} & 82.7\textcolor{gray}{\tiny$\pm$0.7} & \textbf{95.2}\textcolor{gray}{\tiny$\pm$0.2} & 79.7\textcolor{gray}{\tiny$\pm$1.0} & \textbf{60.7}\textcolor{gray}{\tiny$\pm$0.8} & \textbf{89.8}\textcolor{gray}{\tiny$\pm$0.5} & \textbf{65.9}\textcolor{gray}{\tiny$\pm$1.2} & \textbf{75.9}                      \\ 
\hline\hline
\multicolumn{11}{c}{\rule{0pt}{10pt}\fontsize{8.5}{10}\selectfont  \textbf{Data ratio: 1\%}}                                                                                                                  \\ 
\hline
\multirow{2}{*}{Method} & \multicolumn{3}{c}{PTB-XL}     & \multicolumn{3}{c}{Chapman}    & \multicolumn{3}{c}{CPSC2018}   & \multirow{2}{*}{Overall}  \\ 
\cline{2-10}
                        & F1       & AUROC    & ACC      & F1       & AUROC    & ACC      & F1       & AUROC    & ACC      &                           \\ 
\hline
Random                  & 43.8\textcolor{gray}{\tiny$\pm$1.3} & 76.0\textcolor{gray}{\tiny$\pm$1.0} & 61.5\textcolor{gray}{\tiny$\pm$1.8} & 71.0\textcolor{gray}{\tiny$\pm$2.1} & 89.1\textcolor{gray}{\tiny$\pm$1.4} & 71.1\textcolor{gray}{\tiny$\pm$1.8} & 35.6\textcolor{gray}{\tiny$\pm$1.6} & 75.9\textcolor{gray}{\tiny$\pm$0.6} & 43.4\textcolor{gray}{\tiny$\pm$1.5} & 63.0                        \\
\hline
MOCO                    & 43.0\textcolor{gray}{\tiny$\pm$0.8} & 74.6\textcolor{gray}{\tiny$\pm$1.6} & 62.6\textcolor{gray}{\tiny$\pm$0.8} & 75.4\textcolor{gray}{\tiny$\pm$0.8} & 90.3\textcolor{gray}{\tiny$\pm$0.6} & 73.0\textcolor{gray}{\tiny$\pm$1.1} & 38.8\textcolor{gray}{\tiny$\pm$1.3} & 77.9\textcolor{gray}{\tiny$\pm$0.8} & 46.9\textcolor{gray}{\tiny$\pm$1.3} & 64.7                      \\
BYOL                    & 42.5\textcolor{gray}{\tiny$\pm$0.6} & 75.5\textcolor{gray}{\tiny$\pm$0.8} & 65.1\textcolor{gray}{\tiny$\pm$1.0} & 79.3\textcolor{gray}{\tiny$\pm$0.9} & 93.3\textcolor{gray}{\tiny$\pm$0.4} & 77.6\textcolor{gray}{\tiny$\pm$1.6} & 37.2\textcolor{gray}{\tiny$\pm$0.8} & 76.3\textcolor{gray}{\tiny$\pm$0.4} & 46.3\textcolor{gray}{\tiny$\pm$0.3} & 65.9                      \\
CLOCS                   & 42.9\textcolor{gray}{\tiny$\pm$0.4} & 74.9\textcolor{gray}{\tiny$\pm$0.9} & 62.4\textcolor{gray}{\tiny$\pm$0.8} & 76.9\textcolor{gray}{\tiny$\pm$0.6} & 91.8\textcolor{gray}{\tiny$\pm$0.4} & 74.8\textcolor{gray}{\tiny$\pm$0.9} & 39.7\textcolor{gray}{\tiny$\pm$1.0} & 78.1\textcolor{gray}{\tiny$\pm$0.3} & 46.0\textcolor{gray}{\tiny$\pm$1.2} & 65.3                          \\
PCLR                    & 45.6\textcolor{gray}{\tiny$\pm$0.6} & 76.1\textcolor{gray}{\tiny$\pm$0.6} & 64.8\textcolor{gray}{\tiny$\pm$0.3} & 61.3\textcolor{gray}{\tiny$\pm$2.0} & 84.1\textcolor{gray}{\tiny$\pm$0.9} & 61.6\textcolor{gray}{\tiny$\pm$2.3} & 34.4\textcolor{gray}{\tiny$\pm$0.8} & 78.3\textcolor{gray}{\tiny$\pm$0.7} & 44.5\textcolor{gray}{\tiny$\pm$0.2} & 61.2                      \\
COMET                   & 32.2\textcolor{gray}{\tiny$\pm$1.3} & 66.7\textcolor{gray}{\tiny$\pm$1.2} & 53.8\textcolor{gray}{\tiny$\pm$1.1} & 78.5\textcolor{gray}{\tiny$\pm$0.7} & 91.7\textcolor{gray}{\tiny$\pm$0.6} & 78.4\textcolor{gray}{\tiny$\pm$1.9} & 26.8\textcolor{gray}{\tiny$\pm$1.5} & 73.1\textcolor{gray}{\tiny$\pm$1.0} & 38.6\textcolor{gray}{\tiny$\pm$2.1} & 60.0                        \\
ETP                     & \textbf{47.3}\textcolor{gray}{\tiny$\pm$0.3} & 76.8\textcolor{gray}{\tiny$\pm$0.4} & \textbf{67.1}\textcolor{gray}{\tiny$\pm$0.4} & 81.6\textcolor{gray}{\tiny$\pm$0.5} & 94.3\textcolor{gray}{\tiny$\pm$0.3} & 79.4\textcolor{gray}{\tiny$\pm$0.6} & 41.3\textcolor{gray}{\tiny$\pm$2.5} & \textbf{80.1}\textcolor{gray}{\tiny$\pm$1.5} & 49.1\textcolor{gray}{\tiny$\pm$1.2} & 68.6                      \\ 
\hline
\textbf{Ours}                   & 46.8\textcolor{gray}{\tiny$\pm$1.1} & \textbf{77.0}\textcolor{gray}{\tiny$\pm$0.6} & 65.1\textcolor{gray}{\tiny$\pm$0.6} & \textbf{82.9}\textcolor{gray}{\tiny$\pm$0.7} & \textbf{94.9}\textcolor{gray}{\tiny$\pm$0.3} & \textbf{82.1}\textcolor{gray}{\tiny$\pm$1.0} & \textbf{41.6}\textcolor{gray}{\tiny$\pm$2.4} & 78.9\textcolor{gray}{\tiny$\pm$0.7} & \textbf{50.0}\textcolor{gray}{\tiny$\pm$1.6} & \textbf{68.8}                      \\
\hline\hline
\end{tabular}}
\end{spacing}
\end{table*}

\subsection{Experimental Setting}

We evaluate our methods on four public ECG datasets in comparison with four baselines.
Specifically, we investigate ECG classification tasks on various cardiac rhythms including cardiac arrhythmia detection and myocardial infarction detection, etc.
Our experiments follow the one-to-many fine-tuning setup \cite{tfc}: after pretraining, we append a new classification head to the encoder and train both of them on other datasets, using only a small portion of training data.
Our aim is to assess the generalization of the inductive bias introduced by pretraining, even under a data scarcity scenario.

\paragraph{Pretraining datasets.}
We use \textbf{MIMIC-IV-ECG} \cite{mimic, physionet} as the pretraining dataset, which contains approximately 800,000 diagnostic electrocardiograms across nearly 160,000 unique patients.
These diagnostic ECGs use 12 leads and are 10 seconds in length.
For efficiency, we only pick a subset of the dataset (about 16,000 patients).

\paragraph{Finetuning datasets.}
We transfer the pretrained model to 3 downstream ECG datasets:
(1) \textbf{PTB-XL} \cite{ptbxl, physionet} contains 21,799 clinical 12-lead ECG records from 18,869 patients of 10 seconds length alongside 5 different classes.
(2) \textbf{Chapman} \cite{chapman} contains 10,646 12-lead ECG records alongside 11 different classes. We group these classes into 4 major classes following the official suggestion \cite{chapman}.
(3) \textbf{CPSC2018} \cite{cpsc2018} contains 6,877 12-lead ECG records alongside 9 classes.

We preprocess all datasets following \cite{comet} to produce equal length trials and split each dataset into training, validation, and test sets by 80, 10, 10 in a patient-independent way \cite{comet} .

\paragraph{Baselines.}

We compare with 6 methods with different design concept:  MoCo \cite{moco} BYOL \cite{byol}, CLOCS \cite{clocs}, PCLR \cite{pclr}, COMET \cite{comet}, ETP \cite{etp}.
Among these approaches, MoCo and BYOL, originating from computer vision, have served as foundational techniques in various works on ECG \cite{clecg, jepa}.
We employ the same data augmentations for these methods as we do for our own.
CLOCS, PCLR, and COMET leverage the patient contrast mechanism to enhance the learned representations.
Additionally, ETP utilizes BioClinicalBERT \cite{bioclinicalbert} and textual statements describing the ECG for cross-modal pretraining, which necessitates an additional pretrained model and substantial effort for statement annotation.
For a fair comparison, we unify the encoder and the number of training epochs across all methods and adopt the same hyper-parameters as reported in their original papers.
We utilize the open source code of each baseline \footnote{Except for ETP and PCLR, for which official implementations are unavailable; we carefully implement these methods based on the descriptions provided in their respective papers.} and adapt them to ensure consistency with our experimental setup.

\paragraph{Metrics.}
We report accuracy, F1 score (macro-averaged), AUROC (macro-averaged) for all experiments. In addition, to measure the comprehensive performance of the pre-trained model on different datasets, we also introduce an additional metric named "Overall", which represents the average of all metrics across all datasets.

\paragraph{Implementation Details.}
Following previous works, the encoder composes an input projection layer for projection before masking and a dilated CNN \cite{tcn, ts2vec, comet, timesiam}.
It consists of 10 hidden blocks, each following the order "GELU -> DilatedConv -> GELU
-> DilatedConv." A residual connection is applied between the beginning and end of each block. The dilation factor of the convolution in the i-th block is set to 2i. Each hidden dimension of the dilated convolution is set to 64, and the kernel size is set to 3. The output dimension of encoder $K$ is fixed at 320. 

For both the projection and prediction heads, we employ MLPs with three and two layers, using ReLU activations after the hidden layers and batch normalization (BN) after all layers.
For the fine-tuning of the classification head, we employ a two-layer MLP architecture, incorporating batch normalization (BN) and ReLU activation after each hidden layer, and applying dropout after the output layer.

We conduct all experiments using five random seeds (41–45). For each evaluation metric, we report the mean and standard deviation across these five random seeds. All experiments run on a single NVIDIA RTX 4090 GPU.
The optimizer used was AdamW with a warmup learning rate strategy.
For the pretraining, We train for 100 epochs and set the learning rate as 0.001, and $\tau=0.1$, $m=0.999$, $M=16384$. We set the basic batch size to 256, and the entire pretraining takes approximately 1.5 hours. For the finetuning, we train for 50 epochs and set the learning rate as 0.0001. We set the basic batch size to 256, and the entire finetuning across all the downstream datasets takes approximately 1 hour.
% device, optimizer, lr...

% \subsection{Results on Linear Evaluation}

\subsection{Experiment Results}
\label{sec:result}

For each downstream dataset, we evaluate model performance using three levels of labeled training data: 30\%, 10\%, 1\%.
The experimental results are summarized in Table~\ref{tab:result}.

Overall, PMQ achieves the best performance in 21 out of 27 evaluated metrics across the three datasets.
In the remaining five metrics, it ranks second, with performance closely comparable to the best-performing baselines.
Notably, PMQ consistently demonstrates superior overall performance across all data availability settings.
With 30\% of labeled data, PMQ surpasses the strongest baseline, BYOL, by an average margin of 1.5\% across the three datasets.

More importantly, under conditions of severe data scarcity—when only 10\% or 1\% of the training data is available, PMQ exhibits greater robustness compared to other baselines.
It outperforms the best baseline in this setting, PCLR, by an average of 1.7\% across the datasets.
PMQ also achieves results comparable to or better than ETP, which leverages a large language model and auxiliary ECG textual reports for representation learning.
In contrast, our method relies solely on raw ECG data without the aid of pretrained models or external supervision, making it a more efficient and practical approach to learning effective representations.

Aside from ETP, for the few cases where PMQ ranks second to COMET, particularly on the Chapman dataset, we hypothesize that COMET’s advantage arises from its use of contrastive blocks at both the sample and instance levels, which help capture fine-grained features.
This is supported by the ablation studies reported in the original COMET paper.
The relatively lower performance of PMQ in these specific cases may be attributed to the simplicity of its data augmentation strategy, which might limit its ability to exploit low-level instance-specific features that are useful in certain downstream tasks, we will discuss later in Section~\ref{sec:ablation}.
Nevertheless, by fully leveraging patient-level information, PMQ achieves stronger overall results while maintaining a significantly lower computational cost than COMET, whose hierarchical structure entails more intensive computation costs.

We also validate the effectiveness of the Patient Memory Queue (PMQ) in comparison to MoCo. This demonstrates that enhancing a model through patient-level context requires not only an increase in negative samples, as achieved by MoCo's memory queue, but also in positive samples via the patient memory queue.

\begin{table}[h!]
\centering
\begin{spacing}{1.1} % 调整行距
\caption{\textbf{Ablation result.} Only the F1 score is reported. The All at the bottom line indicates original PMQ.}
\label{tab:ablation}
\resizebox{\linewidth}{!}{
\begin{tabular}{ccccc} 
\hline\hline
\multirow{2}{*}{} & \multicolumn{4}{c}{\rule{0pt}{10pt}\fontsize{8.5}{10}\selectfont \textbf{Data ratio: 30\%}}                                       \\ 
\cline{2-5}
                  & ~PTB-XL~          & Chapman           & CPSC2018          & Overall        \\ 
\hline
w/o mask\_t       & 55.3\textcolor{gray}{\tiny$\pm$0.3}          & 84.3\textcolor{gray}{\tiny$\pm$0.7}          & 62.0\textcolor{gray}{\tiny$\pm$1.2}          & 67.2           \\
w/o mask\_f       & 55.9\textcolor{gray}{\tiny$\pm$0.7}          & 88.3\textcolor{gray}{\tiny$\pm$0.5}          & 62.1\textcolor{gray}{\tiny$\pm$1.2}          & 68.8           \\
w/o neighbor      & 56.0\textcolor{gray}{\tiny$\pm$0.8}          & 86.6\textcolor{gray}{\tiny$\pm$1.0}          & 59.7\textcolor{gray}{\tiny$\pm$1.1}          & 67.4           \\
w/o queue         & 55.3\textcolor{gray}{\tiny$\pm$0.3}          & 85.7\textcolor{gray}{\tiny$\pm$0.6}          & \textbf{64.4}\textcolor{gray}{\tiny$\pm$0.7} & 68.5           \\ 
\hline
\textbf{All}      & \textbf{56.1}\textcolor{gray}{\tiny$\pm$0.1} & \textbf{86.8}\textcolor{gray}{\tiny$\pm$0.9} & 64.2\textcolor{gray}{\tiny$\pm$0.9}          & \textbf{69.0}  \\ 
\hline\hline
\multirow{2}{*}{} & \multicolumn{4}{c}{\rule{0pt}{10pt}\fontsize{8.5}{10}\selectfont \textbf{Data ratio: 10\%}}                                       \\ 
\cline{2-5}
                  & ~PTB-XL~          & Chapman           & CPSC2018          & Overall        \\ 
\hline
w/o mask\_t       & 52.1\textcolor{gray}{\tiny$\pm$0.6}          & 79.4\textcolor{gray}{\tiny$\pm$1.4}          & 58.4\textcolor{gray}{\tiny$\pm$0.8}          & 63.3           \\
w/o mask\_f       & 52.9\textcolor{gray}{\tiny$\pm$1.2}          & \textbf{85.0}\textcolor{gray}{\tiny$\pm$1.3} & 58.1\textcolor{gray}{\tiny$\pm$2.2}          & 65.3           \\
w/o neighbor      & 51.8\textcolor{gray}{\tiny$\pm$1.0}          & 80.8\textcolor{gray}{\tiny$\pm$0.9}          & 55.4\textcolor{gray}{\tiny$\pm$1.0}          & 62.7           \\
w/o queue         & 52.2\textcolor{gray}{\tiny$\pm$0.4}          & 81.9\textcolor{gray}{\tiny$\pm$0.6}          & 59.6\textcolor{gray}{\tiny$\pm$1.6}          & 64.6           \\ 
\hline
\textbf{All}      & \textbf{53.6}\textcolor{gray}{\tiny$\pm$0.3} & 82.7\textcolor{gray}{\tiny$\pm$0.7}          & \textbf{60.7}\textcolor{gray}{\tiny$\pm$0.8} & \textbf{65.7}  \\ 
\hline\hline
\multirow{2}{*}{} & \multicolumn{4}{c}{\rule{0pt}{10pt}\fontsize{8.5}{10}\selectfont \textbf{Data ratio: 1\%}}                                      \\ 
\cline{2-5}
                  & ~PTB-XL~          & Chapman           & CPSC2018          & Overall        \\ 
\hline
w/o mask\_t       & \textbf{47.0}\textcolor{gray}{\tiny$\pm$1.4} & 77.1\textcolor{gray}{\tiny$\pm$1.2}          & 40.2\textcolor{gray}{\tiny$\pm$1.3}          & 54.8           \\
w/o mask\_f       & 45.1\textcolor{gray}{\tiny$\pm$1.6}          & 81.3\textcolor{gray}{\tiny$\pm$1.1}          & 41.4\textcolor{gray}{\tiny$\pm$0.5}          & 55.9           \\
w/o neighbor      & 43.7\textcolor{gray}{\tiny$\pm$1.6}          & 71.0\textcolor{gray}{\tiny$\pm$2.1}          & 35.6\textcolor{gray}{\tiny$\pm$1.6}          & 50.1           \\
w/o queue         & 44.3\textcolor{gray}{\tiny$\pm$0.9}          & 79.7\textcolor{gray}{\tiny$\pm$0.3}          & 39.7\textcolor{gray}{\tiny$\pm$0.6}          & 54.6           \\ 
\hline
\textbf{All}      & 46.8\textcolor{gray}{\tiny$\pm$1.1} & \textbf{82.9}\textcolor{gray}{\tiny$\pm$0.7} & \textbf{41.6}\textcolor{gray}{\tiny$\pm$2.4} & \textbf{57.1}  \\
\hline\hline
\end{tabular}}
\end{spacing}
\end{table}

\subsection{Ablation Study}
\label{sec:ablation}
In this section, following the experimental setup in the main experiment, we evaluated the contribution of each individual component in our method. The corresponding results are shown in the Table~\ref{tab:ablation}. In the ablation study, we reported the F1 score and the overall score (representing the average performance on all data). Among them, w/o mask\_t, w/o mask\_f, w/o neighbor, and w/o queue respectively mean removing Timestamp Masking, Frequency Masking, Temporal Neighboring, and the patient memory queue.

\paragraph{Effectiveness of patient memory queue}
In general, eliminating the patient memory queue results in a decline in the overall F1 score, especially in situations of severe data scarcity.
This indicates that the patient memory queue enhances the model's robustness against the challenges posed by real-world data scarcity.

We observe that at higher data availability (e.g., 30\%), removing the patient memory queue results in a slight drop in overall score (decrease 0.5).
When the data ratio decreased to 10\%, the overall score of the model further decreased (by 1.1).
This indicates that when the amount of data is sufficient, the model can benefit from the supervision signals brought by labeled data and reduce its dependence on the context of the ECG series mined from the pretrained model.

However, When the data ratio was only 1\%, the model's performance on all datasets decreased significantly, and the overall score decreased by (2.5).
This indicates that when the amount of data is extremely small, the downstream model cannot obtain sufficient effective supervision signals to guide the model's learning, and at this time, it will rely more on the pretrained representation of the model. When PMQ is removed, the model cannot fully explore the context of the ECG series, resulting in a degradation of the pre-trained model's capabilities. This in turn affects the performance of the model on downstream tasks with very few supervision signals.

\begin{figure}[t]
    \centering
    \includegraphics[width=\linewidth, trim=5 8 5 7, clip]{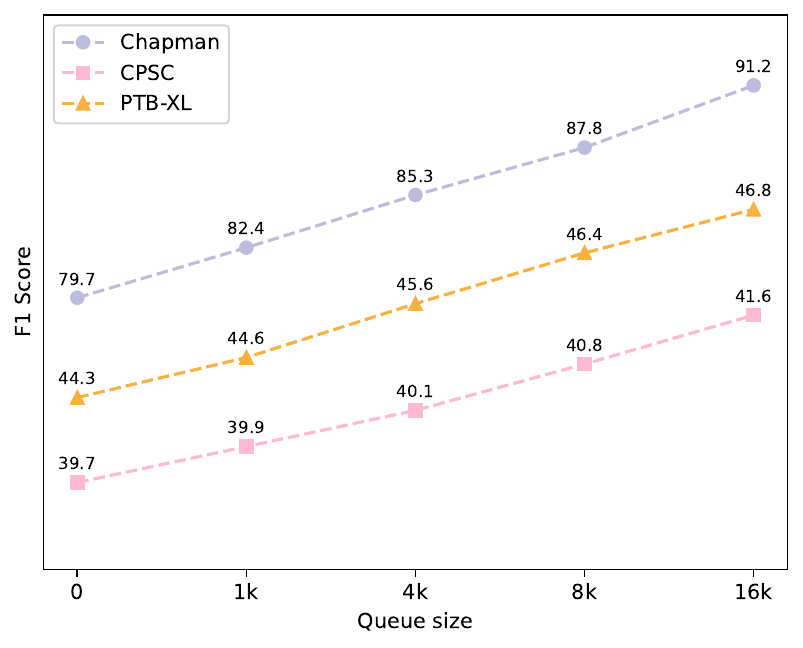}
    \caption{\textbf{Performance of different patient memory queue size.} We report the F1 score with 1\% of all datasets.}
    \label{fig:M}
\end{figure}

\paragraph{Effectiveness of the size M of the patient memory queue}
To further explore the impact of the patient memory queue on the performance of the pre-trained model, we investigated the size of M in the patient memory queue. Since the performance shows the greatest dependence on the pretrained model when there is only 1\% of the data volume in the downstream dataset, we conducted experiments under this setting, as shown in Figure \ref{fig:M}.

From the results of the figure, it can be seen that as M increases, the performance of the model on the three downstream datasets also increases. This verifies the view we put forward at the beginning of the article, that is, by incorporating more inter- and intra-patient sample pairs, the context within the ECG series can be better mined, enabling the ECG pretrained model to provide better representations, thereby enhancing its robustness in downstream tasks.

In addition, when the size of M is 1k, the comprehensive performance of our method is also better than the previous patient contrastive learning methods (such as CLOCS, PCLR, COMET). This also verifies another previous view of ours, that is, the previous patient contrastive learning methods are limited by the batch size, resulting in their inability to fully explore the contextual performance within the ECG series, which limits the upper bound of the model performance.

\paragraph{Effectiveness of data augmentations}
Across all datasets and data ratios, the removal of the neighboring data sampling augmentation leads to the most significant performance degradation.
This highlights the importance of leveraging short-distance temporal correlations, indicating that learning from temporally adjacent samples also plays an important role in representation quality.

Regarding temporal and frequency masking, we generally find that these augmentations improve performance by helping the model capture instance-level features.
However, their impact varies across different datasets and data ratios, suggesting that heterogeneity in data distributions affects augmentation effectiveness.
This variability underscores the potential for developing more unified and robust augmentation strategies that can consistently enhance model performance across diverse settings.

Notably, removing frequency masking results in improved performance on the Chapman dataset with 10\% labeled data, even surpassing the best result previously achieved by COMET (Table~\ref{tab:result}).
This outcome supports our earlier hypothesis discussed in Section~\ref{sec:result}, further validating the design considerations behind PMQ.

\section{Conclusion}

In this paper, we proposed PMQ, a contrastive learning-based ECG pretraining framework enhanced by a dynamic Patient Memory Queue.
Our approach addresses the key limitation of insufficient intra- and inter-patient sample diversity during training by maintaining a large and constantly updated memory queue that preserves patient-level contextual representations.
This enables the model to better exploit patient consistency signals, which are often underutilized in existing patient contrastive learning frameworks.
% We further introduced additional augmentation strategies—temporal neighboring, timestamp masking, and frequency masking—to enrich the diversity of positive and negative pairs and improve instance-level feature learning. 
Through comprehensive evaluations on three public ECG datasets under various labeled data availability settings, our method consistently outperformed both general and patient-level contrastive learning baselines.
Notably, PMQ demonstrated enhanced robustness in low-resource scenarios and achieved competitive or superior performance compared to models that rely on auxiliary ECG textual reports and large language models, despite using only raw ECG signals.
Our results highlight the effectiveness of leveraging patient-level context in a scalable and computationally efficient manner.

\bibliography{mybibfile}

\end{document}